\documentclass[useAMS,usenatbib]{mn2e}

\usepackage{graphicx}

\usepackage{natbib}

\usepackage{times}

\usepackage[T1]{fontenc}
\usepackage{ae,aecompl}

\newcommand{\be}{\begin{displaymath}}
\newcommand{\ee}{\end{displaymath}}
\newcommand{\bea}{\begin{eqnarray}}
\newcommand{\eea}{\end{eqnarray}}

\title[Abundance variations in globular-cluster stars]{The primordial and evolutionary abundance variations in globular-cluster stars: a problem with two unknowns}

\author[P. A. Denissenkov, D. A. VandenBerg, F. D. A. Hartwick et al.]{P. A. 
Denissenkov$^{1,2,5}$\thanks{E-mail: pavelden@uvic.ca.}, D. A. VandenBerg$^{1}$,  
F. D. A. Hartwick$^{1}$, F. Herwig$^{1,2,5}$, \newauthor A. Weiss$^{3}$ and B. Paxton$^{4}$\\
$^{1}$Department of Physics \& Astronomy, University of Victoria,
       P.O.~Box 1700, STN CSC, Victoria, B.C., V8W~2Y2, Canada\\
$^{2}$The Joint Institute for Nuclear Astrophysics, Notre Dame, IN 46556, USA\\
$^{3}$Max-Planck-Institut f\"{u}r Astrophysik, Karl-Schwarzschild-Str. 1, 85741 Garching bei M\"{u}nchen, Germany\\
$^{4}$Kavli Institute for Theoretical Physics and Department of Physics, Kohn Hall, University of California, Santa Barbara, 
CA 93106, USA\\
$^{5}$NuGrid collaboration}
\begin{document}

\date{Accepted 2014 December 31. Received 2014 December 31; in original form 2014 December 31}

\pagerange{\pageref{firstpage}--\pageref{lastpage}} \pubyear{2014}

\maketitle

\label{firstpage}

\begin{abstract}
We demonstrate that among the potential sources of the
primordial abundance variations of the proton-capture elements
in globular-cluster stars proposed so far, such as the hot-bottom burning in massive AGB stars
and H burning in the convective cores of supermassive and fast-rotating massive MS stars, only the supermassive MS stars
with $M > 10^4 M_\odot$ can explain all the observed abundance correlations without any fine-tuning of model parameters.
We use our assumed chemical composition for the pristine gas in M13 (NGC\,6205) and its mixtures with
50\% and 90\% of the material partially processed in H burning in the $6\times 10^4\,M_\odot$ MS model star 
as the initial compositions for the normal,
intermediate and extreme populations of low-mass stars in this globular cluster, as suggested by its O-Na anti-correlation.
We evolve these stars from the zero-age MS to the RGB tip with the thermohaline and parametric prescriptions for the RGB
extra mixing. We find that the $^3$He-driven thermohaline convection cannot explain the evolutionary decline of [C/Fe] in M\,13
RGB stars, which, on the other hand, is well reproduced with the universal values for the mixing depth and rate calibrated using 
the observed decrease of [C/Fe] with $M_V$ in the globular cluster NGC5466 that does not have the primordial abundance variations.
\end{abstract}


\section{Introduction}
\label{sec:intro}

There is a long-standing problem in stellar astrophysics --- understanding the origin of the abundance anomalies of
the proton-capture elements, such as C, N, O, F, Na, Mg, Al, and their isotopes in globular-cluster stars 
\citep[e.g.,][]{kraft:79,kraft:94,gratton:12}. The fact that the anomalous abundance variations display clear anti-correlations
between C and N, O and Na, Na and F, O and Al, as well as Al and Mg is unanimously interpreted as a strong evidence for
all of them to have been produced in hydrogen burning at a sufficiently high temperature, so that reactions of
the NeNa and MgAl cycles were able to compete with the CNO cycle \citep{denissenkov:98,prantzos:07}. 
Given that some of these anti-correlations are
found in low-mass main-sequence (MS) stars in the present-day globular clusters (GCs) \citep{briley:96,cannon:98,gratton:01,
briley:02,briley:04}, the required high-temperature
H burning must have occurred in their more massive siblings in the past. Because anti-correlations of the same magnitude
are observed in a same GC both in MS and red giant stars \citep{gratton:01,dobrovolskas:14}, 
the latter possessing deep convective envelopes, 
quite significant fractions of material lost by the massive stars that had produced those anti-correlations must have been mixed
with the pristine gas in the young GC before the low-mass stars formed out of that mixture.

Three types of H-burning in stars have been proposed as possible sources of the primordial abundance variations of p-capture 
elements in GCs: hot-bottom burning in massive asymptotic giant branch (AGB) stars \citep{dantona:83}, H burning
in convective cores of rapidly rotating massive MS stars \citep{decressin:07} and, more recently, core H burning in
supermassive MS stars with masses $M > 10^4\,M_\odot$ \citep{denissenkov:14}. Fast rotation with a nearly break-up
velocity in the second case plays a twofold role: firstly, it drives rotation-induced mixing in the radiative envelope, thus
bringing H-burning products from the convective core to the surface and, secondly, it leads to equatorial mass loss 
with a relatively low velocity caused by the centrifugal force. The second property, like the assumed low-velocity
mass loss by the AGB stars, is required to explain the retention of the mass lost by the massive stars in the shallow potential 
well of the young GC. It is also assumed that the massive AGB and MS stars had migrated to the GC centre, as a result of
dynamical friction, before they deposited
the products of H burning to the GC interstellar medium. The last assumption is usually used to interpret the larger    
fraction of low-mass stars with the stronger abundance anomalies in the cores of some GCs \citep[e.g.][]{milone:12}.

A solution of the problem of the primordial abundance variations in GCs should be divided into two steps. The first step is to find 
the right massive star candidate, such that when we dilute its H-burning yields with the GC pristine gas we get individual 
abundances and correlations between them consistent with all the relevant observational data for GCs. The second step is 
to understand how the required massive stellar objects formed and functioned, how their lost mass was retained and 
mixed with the pristine gas in GCs, and how significant fractions of low-mass stars made out of those gas mixtures survived and 
got distributed among their unpolluted counterparts in GCs by the present day. In the framework of this two-step
solution, the early disc accretion model of \cite{bastian:13a} and the model of massive binaries of \cite{demink:09}
are not considered by us as additional possibilities to explain the origin of the p-capture element abundance anomalies in GCs, 
because either of these models still uses massive MS stars as the H-burning source of those anomalies.
Other details of these models are relevant only to the second step.
However, we think that it is not worth discussing any solution details pertaining to the second step until the first step is 
completed, especially, given that there are a lot of observational data on abundances of p-capture elements and their isotopes in
GC stars to constrain the solution on the first step, whereas there are no direct observational data on
the formation of the first generation stars in GCs. 

In this paper, we conclude that the massive AGB and MS stars 
are not the best candidates for the origin of the primordial abundance variations in GCs, because they
fail to reproduce the correlations between the abundances of Al and Mg isotopes, that have recently been reported by 
\cite{dacosta:13} and now include stars from 5 GCs. 
We demonstrate that this failure is a consequence of temperatures of H burning in these objects that are
either too high (in the case of AGB stars) or too low (in the case of massive MS stars).
On the other hand, the hypothetical supermassive MS stars with $M > 10^4\,M_\odot$ have the right temperature to nicely 
reproduce not only the Mg-Al anti-correlation, but also all the other observed abundance anomalies of the p-capture elements in GCs,
including enhanced He abundances \citep{denissenkov:14}. 

The primordial variations of the C and N abundances and $^{12}$C/$^{13}$C isotopic ratio in GCs are obscured by their evolutionary 
changes that occur in low-mass stars, both in GCs and in the field, on the upper red giant branch (RGB) above the bump luminosity. 
These changes are caused by some extra mixing that operates in radiative zones of RGB stars between the H-burning shell (HBS) 
and the bottom of the convective envelope (BCE). It results in the decreasing surface C abundance and $^{12}$C/$^{13}$C ratio and 
increasing N abundance when the star climbs the upper RGB and its luminosity increases. At the bump luminosity,
the HBS, advancing in mass, erases a mean molecular weight ($\mu$) discontinuity left behind by the BCE at the end of the first 
dredge-up.  This discontinuity probably prevents extra mixing from reaching the HBS on the lower RGB. Above the bump luminosity,
the $\mu$-profile in the radiative zone is uniform everywhere, except the vicinity of the HBS, where
the reaction $^3$He($^3$He,2p)$^4$He produces its local depression of the order of $\Delta\mu\sim 10^{-4}$
\citep{eggleton:06}. This $\mu$-depression should drive thermohaline mixing that was proposed for the role of RGB extra mixing
by \cite{charbonnel:07} who actually assumed that its associated fluid parcels 
(``salt fingers'') had a ratio of their vertical length 
to horizontal diameter $a = l/d\approx 6.2$. However, numerical simulations of thermohaline convection by
\cite{denissenkov:10} have shown that the aspect ratio of salt fingers in RGB stars is rather $a\la 1$. Given that
the diffusion coefficient for thermohaline convection $D_\mathrm{th}$ is proportional to $a^2$, it turns out to be too inefficient 
for the RGB extra mixing.

In this work, we have chosen M13 (NGC\,6205) as an exemplary instance of GCs with abundance anomalies of p-capture elements,
because it is one of a few GCs that show the most extreme primordial abundance anomalies.
We assume that its low-mass stars had been formed out of mixtures of the pristine gas and a varying
fraction of material lost by a supermassive MS star with $M > 10^4\,M_\odot$, as described by \cite{denissenkov:14}.
Following \cite{johnson:12}, we use the O-Na anti-correlation for M13 stars to subdivide them 
into three populations according to the strength of
their primordial abundance anomalies: a normal (or primordial) population is made of the pristine gas, 
while intermediate and extreme
populations contain, respectively, 50\% and 90\% of material from the supermassive star mixed with the pristine gas.
Then, we allow the low-mass stars belonging to the different populations to evolve from the zero-age MS to the RGB tip. 
The RGB extra mixing is modeled either
using the thermohaline diffusion coefficient (equation 25) from \cite{denissenkov:10}  with the salt-finger aspect ratios 
$a\ga 7$ that provide the most efficient mixing or using the observationally constrained parametric prescription
from \cite{denissenkov:08} and \cite{denissenkov:12} that employs the same mixing depth as in the thermohaline case, i.e.
$\log_{10} (r_\mathrm{mix}/R_\odot) = -1.35$, and diffusion coefficient $D_\mathrm{mix} = \alpha K$, 
where $K$ is the thermal diffusivity
and $\alpha = 0.01$\,--\,$0.1$ is the free parameter. This simple model that focuses on the nucleosynthesis part 
(the first step) of the solution takes into account both the primordial and evolutionary abundance variations of
the p-capture elements in GC stars. We compare its predictions with the relevant observational data not only for M13 but
also for other GCs.

\begin{figure*}
\includegraphics[width=15cm,bb = 99 256 547 603]{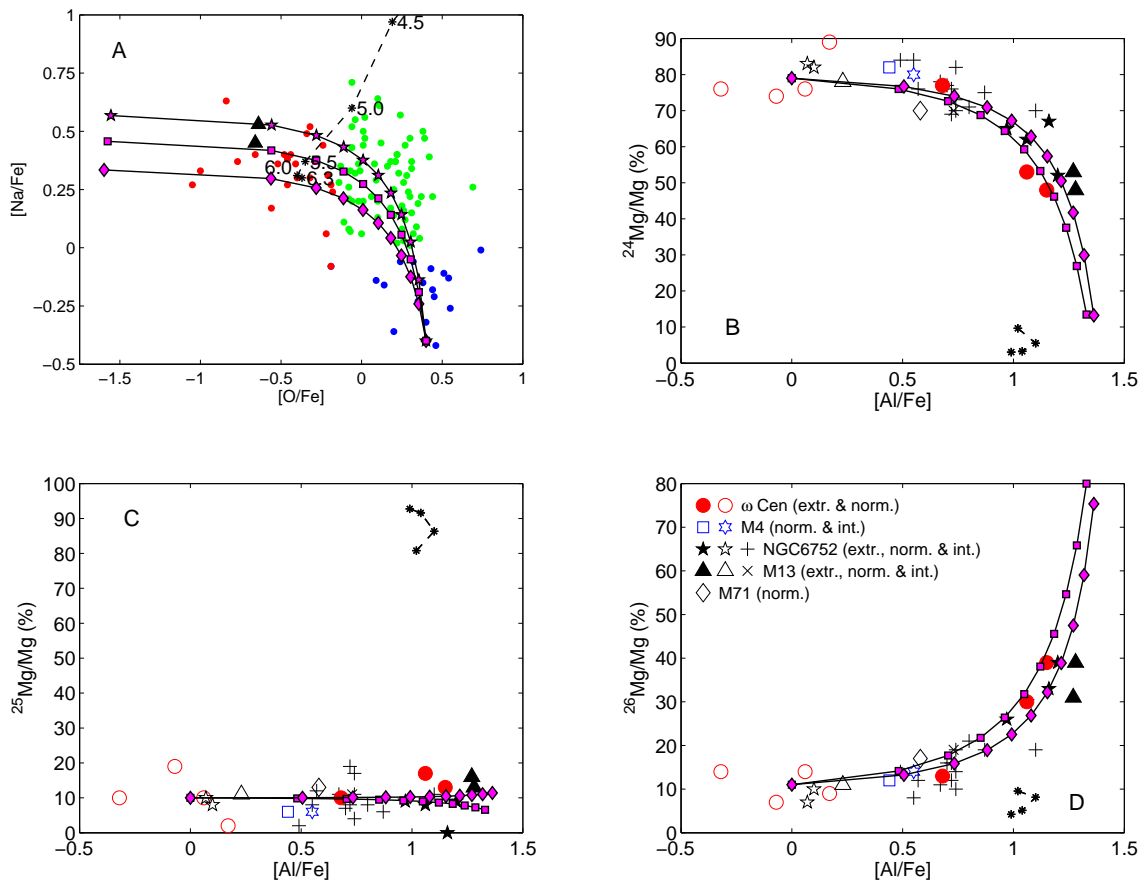}
\caption{\small Panel A: the O-Na anti-correlation for the M\,13 RGB stars (the blue, green and red circles) from 
         \protect\cite{johnson:12} is compared with
         the dilution curves (the magenta star symbols, squares and diamonds connected by the solid black curves) obtained by 
         mixing the abundances from the M\,13 pristine gas (the lower-right ends of the curves) with a varying fraction (from 0\%
         to 100\%) of material from the supermassive MS stars with the masses $5\times 10^4\,M_\odot$ (the star symbols),
         $6\times 10^4\,M_\odot$ (the squares) and $7\times 10^4\,M_\odot$ (the diamonds). The black asterisks connected
         by the dashed line are the theoretical data for the massive AGB stars with the indicated initial masses from
         \protect\cite{ventura:09}. Panels B, C and D: the Al abundances and Mg isotopic ratios for the 33 RGB stars 
         from 5 GCs collected by \protect\cite{dacosta:13} (the large single symbols, as identified in panel D) are compared 
         with the theoretical predictions 
         from the supermassive MS and massive AGB stars, as explained for panel A.
         }
\label{fig:ona_mg}
\end{figure*}

\section{Computational method}
\label{sec:method}

The MS evolution of supermassive stars with $M > 10^4\,M_\odot$ is calculated using the revision 5329 of the stellar evolution 
code of MESA\footnote{\tt http://mesa.sourceforge.net} \citep{paxton:11,paxton:13}, as described by \cite{denissenkov:14}. 
In particular, because the MS stars with $M > 10^4\,M_\odot$  are objects with super-Eddington luminosities, in which 
the radiation pressure dominates over the gas pressure, we use the MLT++ prescription for convection that is recommended 
in MESA for such cases (see section 7.2 in \citealt{paxton:13}). Other input physics data and assumptions in our calculations of
supermassive MS stars are the same that we use to calculate low-mass star models.

The evolution of low-mass stars is computed using the older MESA revision 3251.
The RGB extra mixing is modeled with the thermohaline diffusion coefficient $D_\mathrm{mix} = D_\mathrm{th}$ and
parametric prescription $D_\mathrm{mix} = \alpha K$. Model smoothing parameters were adjusted by \cite{denissenkov:10} 
for the MESA revision 3251 to reproduce the results of his COMSOL high-resolution test simulations of thermohaline mixing
in low-mass RGB stars. Stellar evolution codes from both revisions are run with the same nuclear network that includes 
31 isotopes from $^1$H to $^{28}$Si coupled by 60 reactions of
the pp chains, CNO, NeNa and MgAl cycles. For solar composition, we use the elemental abundances of \cite{grevesse:98}
with the isotopic abundance ratios from \cite{lodders:03}. The chemical composition of the pristine gas in M13 is obtained
from the solar composition using [Fe/H]\footnote{We use the standard spectroscopic notation
[A/B]\,$ = \log_{10}(N(\mathrm{A})/N(\mathrm{B})) - \log_{10}(N_\odot (\mathrm{A})/N_\odot (\mathrm{B}))$,
where $N(\mathrm{A})$ and $N(\mathrm{B})$ are number densities of the nuclides A and B.}\,$= -1.53$ for the metallicity of 
M13 as a scaling factor and [$\alpha$/Fe]\,$= +0.4$ 
for the abundances of $\alpha$-elements ($^{16}$O, $^{20}$Ne, $^{24}$Mg etc.). We also assume [Na/Fe]\,$= -0.4$ and the solar
Mg isotopic ratios in the initial composition of the pristine gas because these values are suggested by the observational
data (e.g., Fig.~\ref{fig:ona_mg}). This initial chemical composition is slightly different from that used by 
\cite{denissenkov:14}, therefore we have re-calculated our supermassive MS models.
With the MESA {\tt kap} pre-processor we have generated opacity tables appropriate for
our initial composition, i.e. the ones based on the \cite{grevesse:98} solar abundances with [$\alpha$/Fe]\,$= +0.4$, and
employed them for all of our mixtures. The pre-processor uses the corresponding OPAL opacity tables \citep{iglesias:93,iglesias:96} 
and low-temperature molecular opacities of \cite{ferguson:05} as input data. For the convective mixing length, 
we have chosen the MESA solar calibrated parameter $\alpha_\mathrm{MLT} = 1.92$ in the \cite{henyey:65} MLT prescription. 
Atmospheric boundary conditions are calculated in the approximation of \cite{swamy:66}.

For M13 isochrone calculations, we have used the Victoria stellar evolution code \citep{vandenberg:12}, because it treats 
both atomic diffusion and its counteracting turbulent mixing, whereas MESA code does not include 
the latter, which leads to an excessive depletion of the surface He abundance on the MS\footnote{When very close to the same 
physics is assumed, the Victoria and MESA codes predict nearly identical evolutionary tracks and lifetimes for stars of a given 
mass and chemical composition \protect\citep[see][]{vandenberg:12}.}. Therefore, in our MESA calculations of
the evolution of low-mass stars we neglect atomic diffusion. This increases the effective temperature at the MS turn-off,
but does not lead to important differences, except for Li, in the evolution of surface composition on the RGB 
as compared to the Victoria
models, because the first dredge-up erases most of the surface abundance changes produced by atomic diffusion on the MS.

\begin{figure}
\includegraphics[width=8cm,bb = 105 242 471 527]{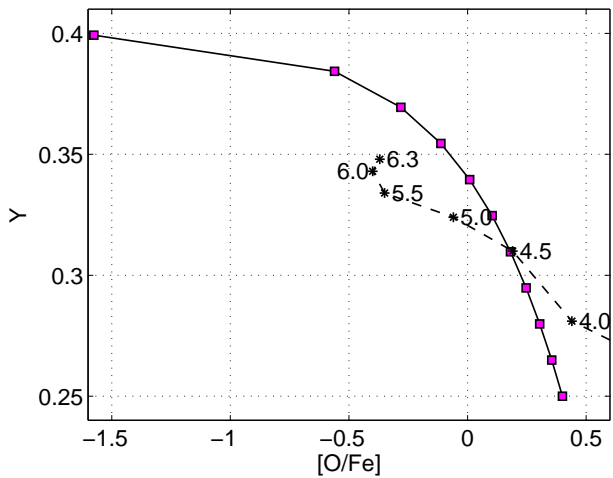}
\caption{\small The same theoretical plot as in Fig.~\ref{fig:ona_mg}A, but for the He mass fraction $Y$. The dilution
         curves for the different supermassive MS star models are overlaying one another.
         }
\label{fig:ofey}
\end{figure}

\section{Three populations of low-mass stars in M\,13}
\label{sec:pops}

Following \cite{denissenkov:14}, we calculate the evolution of supermassive MS stars with $M > 10^4\,M_\odot$ only until
the He mass fraction at the surface and, because these stars are fully convective, also at the centre has reached the value
$Y = 0.40$, which is close to the maximum He abundances reported in the present-day GC stars \citep{pasquini:11,king:12}.
The corresponding ages of the supermassive stars are less than $10^5$ years. Their initial chemical composition is assumed
to be that of the M13 pristine gas, which is equivalent to the composition of the M13 normal population in
Table~\ref{tab:tab1}.

We explicitly require that the evolution of supermassive stars is terminated early on the MS, because
the observed abundance patterns in GCs are characteristic of incomplete H burning
\citep{denissenkov:98,prantzos:07}. As proposed by \cite{denissenkov:14}, the most likely cause of such an early death of 
supermassive stars could be their fragmentation by a diffusive mode of the Jeans instability \citep{thompson:08}, 
which does not depend on a spatial scale and grows on the right time scale of the order of a few $10^5$ years.

\begin{table*}
\caption{Initial mass fractions of isotopes used in our calculations}
\label{tab:tab1}
\begin{tabular}{clll}
\hline
isotope & norm. pop.$^{a}$ & int. pop.$^{b}$ & extr. pop.$^{c}$ \\
\hline
$^{1}$H & $0.748815$ & $0.674238$ & $0.614576$ \\
$^{3}$He & $6.55411\times 10^{-5}$ & $3.27706\times 10^{-5}$ & $6.55416\times 10^{-6}$ \\
$^{4}$He & $0.250000$ & $0.324641$ & $0.384353$ \\
$^{7}$Li & $1.04586\times 10^{-9}$ & $5.23263\times 10^{-10}$ & $1.05187\times 10^{-10}$ \\
$^{12}$C & $9.50177\times 10^{-5}$ & $5.77948\times 10^{-5}$ & $2.80165\times 10^{-5}$ \\
$^{13}$C & $1.15310\times 10^{-6}$ & $3.50972\times 10^{-6}$ & $5.39501\times 10^{-6}$ \\
$^{14}$N & $2.80540\times 10^{-5}$ & $3.52988\times 10^{-4}$ & $6.12936\times 10^{-4}$ \\
$^{15}$N & $1.10507\times 10^{-7}$ & $6.43548\times 10^{-8}$ & $2.74331\times 10^{-8}$ \\
$^{16}$O & $6.55465\times 10^{-4}$ & $3.31200\times 10^{-4}$ & $7.17876\times 10^{-5}$ \\
$^{17}$O & $1.03384\times 10^{-7}$ & $5.94662\times 10^{-8}$ & $2.43319\times 10^{-8}$ \\
$^{18}$O & $5.88641\times 10^{-7}$ & $2.94323\times 10^{-7}$ & $5.88687\times 10^{-8}$ \\
$^{19}$F & $1.38744\times 10^{-8}$ & $6.94144\times 10^{-9}$ & $1.39506\times 10^{-9}$ \\
$^{20}$Ne & $1.35740\times 10^{-4}$ & $1.31317\times 10^{-4}$ & $1.27779\times 10^{-4}$ \\
$^{21}$Ne & $1.36018\times 10^{-7}$ & $6.96419\times 10^{-8}$ & $1.65412\times 10^{-8}$ \\
$^{22}$Ne & $4.37084\times 10^{-6}$ & $2.21363\times 10^{-6}$ & $4.87857\times 10^{-7}$ \\
$^{23}$Na & $1.16195\times 10^{-6}$ & $4.76080\times 10^{-6}$ & $7.63988\times 10^{-6}$ \\
$^{24}$Mg & $4.37780\times 10^{-5}$ & $2.34966\times 10^{-5}$ & $7.27160\times 10^{-6}$ \\
$^{25}$Mg & $5.77473\times 10^{-6}$ & $3.70320\times 10^{-6}$ & $2.04597\times 10^{-6}$ \\
$^{26}$Mg & $6.60731\times 10^{-6}$ & $1.36472\times 10^{-5}$ & $1.92791\times 10^{-5}$ \\
$^{27}$Al & $2.01755\times 10^{-6}$ & $2.25630\times 10^{-5}$ & $3.89994\times 10^{-5}$ \\
$^{28}$Si & $5.69495\times 10^{-5}$ & $5.86796\times 10^{-5}$ & $6.00638\times 10^{-5}$ \\
$^{56}$Fe & $3.92893\times 10^{-5}$ & $3.92893\times 10^{-5}$ & $3.92893\times 10^{-5}$ \\
$^{58}$Ni & $4.33457\times 10^{-5}$ & $4.33457\times 10^{-5}$ & $4.33457\times 10^{-5}$ \\
$[(\mathrm{C}+\mathrm{N}+\mathrm{O})/\mathrm{H}]$ & $-1.24$ & $-1.19$ & $-1.15$ \\
\hline
\end{tabular}

\medskip

$^{a}$ The heavy-element mass fraction for this and the other two mixtures is $Z\approx 0.0011$.

$^{b}$For the intermediate  population, we assume a mixture of 50\% of the abundances from the normal population 
with 50\% of the abundances from our $6\times 10^4\,M_\odot$ MS star model when its He abundance has increased to $Y=0.40$.

$^{c}$For the extreme population, we assume a mixture of 10\% of the abundances from the normal population 
with 90\% of the abundances from our $6\times 10^4\,M_\odot$ MS star model when its He abundance has increased to $Y=0.40$.

\end{table*}

The filled blue, green and red circles in Fig.~\ref{fig:ona_mg}A represent, respectively, the normal (or primordial), intermediate 
and extreme populations of low-mass RGB stars in M\,13, according to the selection criteria used by \cite{johnson:12}. Together, 
they form the O-Na anti-correlation which, like the other correlations between the p-capture elements and their isotopes, is usually
interpreted as a result of mixing of the pristine gas with different fractions of material lost by massive stars that
had taken place in the young GC, before those low-mass stars formed. 
In this interpretation, one end of a correlation, e.g. the lower-right end
of the O-Na anti-correlation, gives abundances in the pristine composition ([O/Fe]\,$\approx +0.4$ and
[Na/Fe]\,$\approx -0.4$ for M13), while the other end points towards abundances characteristic of the polluting star
([O/Fe]\,$\la -1$ and [Na/Fe]\,$\approx +0.4$ for M\,13).

The filled magenta star symbol, square and diamond at the left ends of solid black curves in Fig.~\ref{fig:ona_mg}A
give the O and Na abundances in the MS stars with the masses $5\times 10^4\,M_\odot$, $6\times 10^4\,M_\odot$ and
$7\times 10^4\,M_\odot$ at $Y=0.40$. The same symbols at other locations along the solid black curves show the results of 
these final abundances having been mixed with 10\%, 20\%,\ldots, 90\% and 100\%  of the O and Na abundances from the pristine gas.
As the abundance mixtures representative for the normal, intermediate and extreme populations of stars in M13, we choose
those with 0\%, 50\% and 90\% of material from the MS star with $M = 6\times 10^4 M_\odot$ 
(the first, sixth and tenth magenta squares on the middle solid black
curve counting from its lower-right end). Their corresponding initial isotopic mass fractions are given in Table~\ref{tab:tab1}.
We make such a discrete choice of the initial chemical composition only for simplicity, while keeping in mind that, 
actually, there are no sharp boundaries between the three populations in Fig.~\ref{fig:ona_mg}A.

The different single symbols in Figs.~\ref{fig:ona_mg}B, \ref{fig:ona_mg}C and \ref{fig:ona_mg}D form the discernable dependences of
the Mg isotopic ratios on the Al abundance. They represent observational data for 33 RGB stars from 5 GCs, including
M13, that have recently been collected by \cite{dacosta:13}. The filled magenta squares and diamonds connected by the solid
black curves show the Al and Mg isotopic abundances for the same supermassive star models and mixtures as
in Fig.~\ref{fig:ona_mg}A. The four panels in Fig.~\ref{fig:ona_mg} illustrate the fact, already discussed by
\cite{denissenkov:14}, that the supermassive stars with $M > 10^4\,M_\odot$ reproduce all the primordial 
abundance variations of p-capture elements in GCs surprisingly well. 
The two filled black triangles in Fig.~\ref{fig:ona_mg} are M13 stars that 
belong to the extreme population. From their locations in the four panels, we conclude that all the six abundances in these stars 
are consistent with the supermassive star hypothesis.

\begin{figure}
\includegraphics[width=8cm,bb = 102 247 482 527]{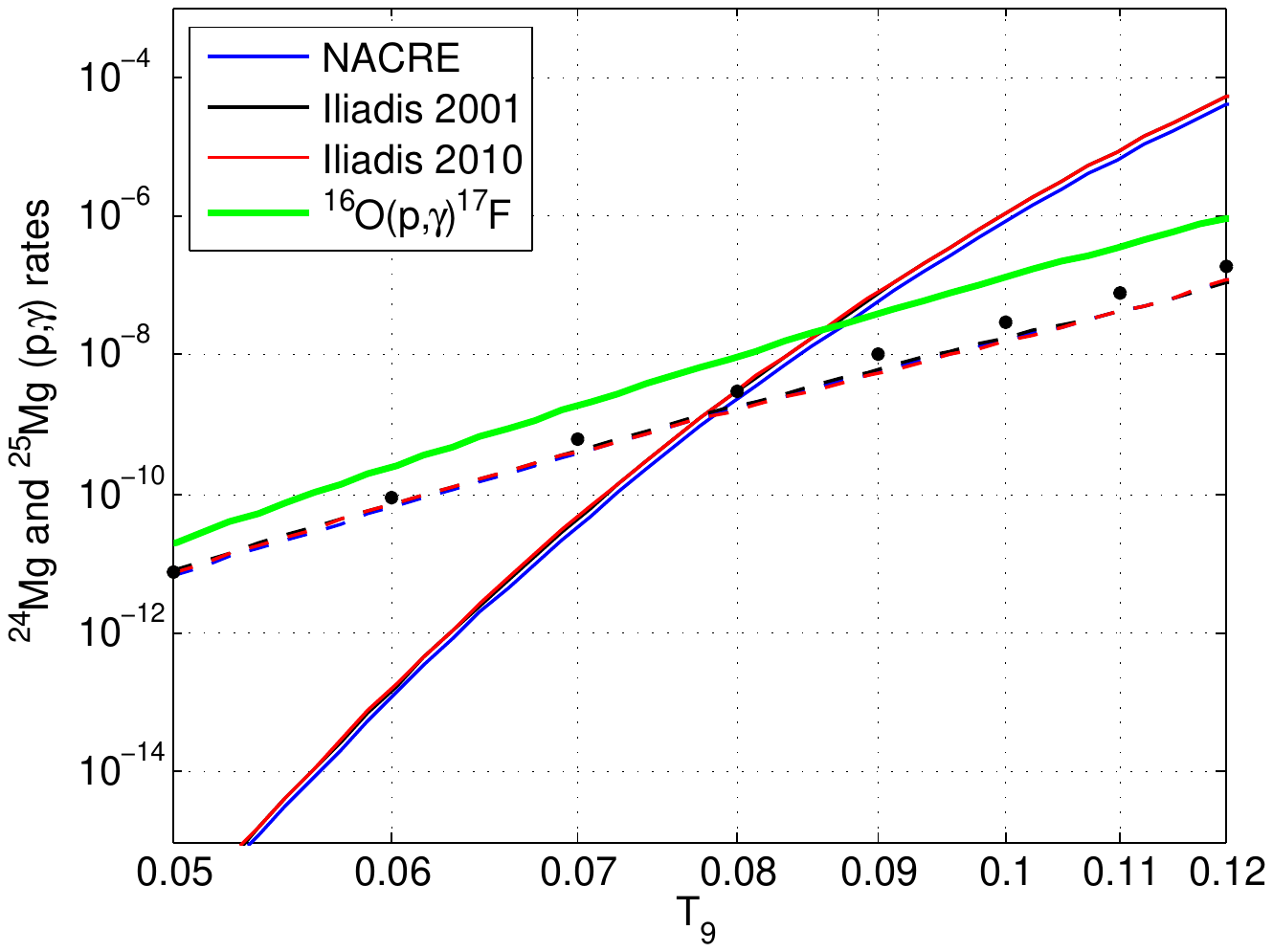}
\caption{\small The (p,$\gamma$) reaction rates (cm$^3$s$^{-1}$mol$^{-1}$) for $^{24}$Mg (the solid blue, black and red curves,
         the last two overlaying one another) 
         and $^{25}$Mg (the dashed curves of the same colors) from the different compilations indicated in the legend box
         (including \protect\cite{iliadis:01} and NACRE data from \protect\cite{angulo:99}), 
         as well as for $^{16}$O (the green curve) from \protect\cite{iliadis:10}.
         These data have been found using a Web interface to the JINA REACLIB default
         library \protect\citep{cyburt:10}. The black circles are the most recent data for the reaction
         $^{25}$Mg(p,$\gamma)^{26}$Al from \protect\cite{straniero:13}. 
         }
\label{fig:mgpg_rates}
\end{figure}

\cite{denissenkov:14} have noted that the success of the supermassive star models in the reproducing           
of the primordial abundance variations of the p-capture elements in GCs,
including the Mg-Al anti-correlation, is not surprising, because these models have central temperatures in the right range
for this, $74\times 10^4\ \mbox{K}\la T_\mathrm{c}\la 78\times 10^6$ K, as was first shown by \cite{denissenkov:98} 
in their ``black box'' solution and, later, independently confirmed by \cite{prantzos:07}. However, in both of the cited papers
the H burning was considered to take place at a constant temperature and, as a result, the required final abundances were reached
when less than 5\% of H was consumed, which would not be sufficient to explain the He enhancements of up to $Y\approx 0.4$ measured 
in some GCs. In the fully convective supermassive MS stars, as much as 20\% of H can be transformed into He, thus changing $Y$ 
from its initial value 0.25 to 0.4, by the moment when the p-capture elements and their isotopes still have the required abundances.
The filled magenta squares in Fig.~\ref{fig:ofey} show the He and O abundances in the mixtures of the M13 pristine gas with
different fractions of the material from the supermassive MS stars. According to this figure, the normal, intermediate and 
extreme populations of stars in M13 should have $Y=0.25$, $Y=0.32$ and $Y=0.38$, respectively (see Table~\ref{tab:tab1}).
These values agree with the He abundances in the blue horizontal branch stars in the GC NGC2808 measured by
\cite{marino:14}.

\section{Supermassive MS stars with $M > 10^4\,M_\odot$ versus massive MS and AGB stars}
\label{sec:agb}

\subsection{The Al abundance and Mg isotopic ratios}

Hydrogen burning in the convective cores of MS stars with $M \la 10^3\,M_\odot$ that occurs at $T_\mathrm{c}\la 60\times 10^6$ K
as long as $Y < 0.40$ can result only in a marginal depletion of the $^{24}$Mg abundance. 
Therefore, neither the fast-rotating massive
MS stars with $20\,M_\odot\leq M\leq 120\,M_\odot$ \citep{decressin:07}, nor the stars with $M=20\,M_\odot$ 
in close binaries \citep{demink:09}, nor the very massive MS stars with
$M\sim 10^3\,M_\odot$ \citep{sills:10}, all of which have been proposed as the potential sources of the primordial abundance
variations of the p-capture elements in GCs, can actually reproduce the observed patterns between the abundances of 
Al and Mg isotopes in Figs.~\ref{fig:ona_mg}B, \ref{fig:ona_mg}C and \ref{fig:ona_mg}D.

The four asterisks connected by the dashed line in Figs.~\ref{fig:ona_mg}B, \ref{fig:ona_mg}C and \ref{fig:ona_mg}D represent
the theoretical data for the AGB stars with the initial masses $5.0\,M_\odot$, $5.5\,M_\odot$, $6.0\,M_\odot$ and $6.3\,M_\odot$
and heavy-element mass fraction $Z=10^{-3}$, which is close to that of M13 stars, from Table~2 of \cite{ventura:09}.
They are located far away from the observed dependences which, on the other hand, are very well matched by the H-burning yields from
the supermassive stars. Unlike the massive MS stars, the problem with the massive AGB stars is that the hot-bottom burning (HBB) of 
H in their convective envelopes occurs at too high temperatures, $T_\mathrm{HBB} \ga 10^8$ K. In Fig.~\ref{fig:mgpg_rates}, we have 
plotted the (p,$\gamma$) reaction rates as functions of $T_{9}\equiv T/10^9\,\mbox{K}$ for $^{24}$Mg (the solid blue, black and
red curves), $^{25}$Mg (the dashed curves) and $^{16}$O (the green curve)
taken from the most recent experimental data 
compilations that we found using a Web interface\footnote{\tt https://groups.nscl.msu.edu/jina/reaclib/db/} 
to the JINA REACLIB default library \citep{cyburt:10}. This figure shows that at $T_{9} \la 0.06$ the reaction
$^{24}$Mg(p,$\gamma)$ is more than three orders of magnitude slower than $^{16}$O(p,$\gamma)$. This explains why
H burning in massive MS stars is not accompanied by the required depletion of $^{24}$Mg. On the other hand,
during the HBB in the massive AGB stars at $T_{9} \ga 0.1$ the rate of the reaction $^{24}$Mg(p,$\gamma)$ exceeds that of 
$^{16}$O(p,$\gamma)$. This should lead to a faster destruction of the most abundant magnesium isotope $^{24}$Mg than
$^{16}$O, which could be a problem, because [O/Fe] usually exhibits much lower values than [Mg/Fe] in GCs \citep{denissenkov:03},
unless the $^{24}$Mg destruction would lead to a commensurate accumulation of $^{25}$Mg. 
This is exactly what happens in the massive AGB stars,
because the reaction $^{25}$Mg(p,$\gamma)$ at $T_{9}\ga 0.1$ is slower than both the p-captures by $^{24}$Mg and $^{16}$O
(Fig.~\ref{fig:mgpg_rates}). Only at $T_{9}\approx 0.075$, which is close to the central temperatures in supermassive MS stars
with $M > 10^4\,M_\odot$, we do find the right relative rates of the above three reactions, which guarantees that when $^{16}$O is
destroyed, a smaller amount of $^{24}$Mg can also be burned, while the freshly produced $^{25}$Mg will be rapidly     
converted into $^{26}$Mg because its p-capture rate is higher than that of $^{24}$Mg(p,$\gamma)$. This explains why
both the massive MS and AGB stars fail to reproduce the observed (anti-)correlations between the abundances of Al and Mg
isotopes, while the H burning in the supermassive MS stars with $M > 10^4\,M_\odot$ does the work.

The black circles in Fig.~\ref{fig:mgpg_rates} present the new rate for the reaction $^{25}$Mg(p,$\gamma)^{26}$Al from
\cite{straniero:13}, which is approximately two times as large as the older rates in the range of $T_{9}$ characteristic of
the HBB in the massive AGB stars. \cite{ventura:11} estimated that with such the increase of this reaction rate they could obtain 
the [Mg/Fe] depletion and [Al/Fe] enhancement in a better agreement with observations. However, their Mg isotopic ratios in
this case, $^{25}$Mg/Mg\,=\,90\% and $^{26}$Mg/Mg\,=\,5.4\%, as well as the ratios $^{25}$Mg/Mg\,=\,76\% and $^{26}$Mg/Mg\,=\,5.6\%
from the super-AGB star with the initial mass $8\,M_\odot$ are still far away from the observed ones.

\begin{figure}
\includegraphics[width=8cm,bb = 106 244 477 531]{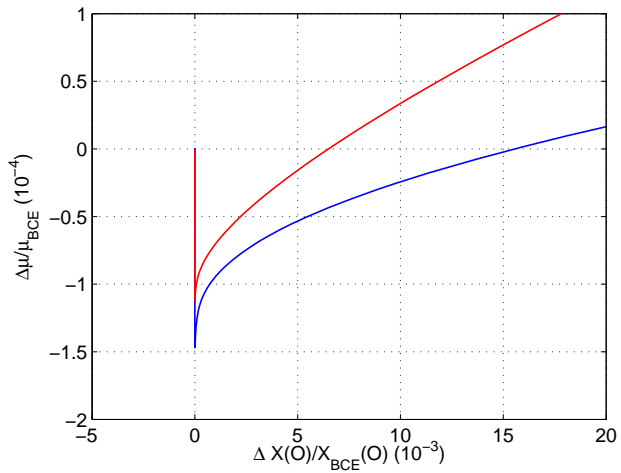}
\caption{\small The relative changes of the mean molecular weight and oxygen abundance in the radiative zones of
         the normal (the blue curve) and extreme, i.e. He-rich, (the red curve) population RGB models with 
         the masses $0.8\,M_\odot$ and
         $0.65\,M_\odot$, respectively, immediately above the bump luminosity. The $\mu$ ratio drops vertically
         in the vicinity of the H-burning shell (HBS), where the reaction $^3$He($^3$He,2p)$^4$He decreases $\mu$ locally, but
         the O abundance has not changed yet. Deeper in the HBS (to the right), the CNO cycle increases both $\mu$ and the relative
         deviation of the O abundance from its value at the bottom of the convective envelope (BCE).
         }
\label{fig:dodmu}
\end{figure}

\subsection{The O-Na anti-correlation}

The minimum value of [O/Fe], that is still accompanied by a relatively high value of [Na/Fe] to fit the O-Na anti-correlation,
obtained in the massive AGB models with the HBB is close to $-0.5$ (see the asterisks connected by the dashed line 
in Figs.~\ref{fig:ona_mg}A and \ref{fig:ofey}). This presents another problem for the massive AGB star pollution hypothesis 
because some stars in M13 
(the red circles to the left of [O/Fe]\,$=-0.5$), as well as stars in a few other GCs, possess much lower O abundances.
To solve this problem, \cite{dantona:07} have proposed that the low-mass stars from the extreme population of GCs experience deeper
extra mixing on the RGB than their counterparts from the normal population because the higher initial He abundance in the former
(Fig.~\ref{fig:ofey}) should reduce the $\mu$-discontinuity that prevents the RGB extra mixing from penetrating deep into the HBS.
However, what really matters when one considers extra mixing in the radiative zone of an RGB star 
is its ability to overcome the restoring Archimedes force that is proportional to $\Delta\mu/\mu_\mathrm{BCE}$, where 
$\Delta\mu = \mu(r) - \mu_\mathrm{BCE}$, provided that $D_\mathrm{mix}\ll K$ which is true for the RGB extra mixing
(see the next section). In Fig.~\ref{fig:dodmu}, we compare the ratios $\Delta\mu/\mu_\mathrm{BCE}$ plotted
as functions of a relative deviation of the local O mass fraction from its value at the BCE (we use the positive difference
$\Delta X(\mbox{O})=X_\mathrm{BCE}(\mbox{O})-X(\mbox{O})$) in our $0.8\,M_\odot$ and $0.65\,M_\odot$ RGB models with the normal and
extreme initial compositions from Table~\ref{tab:tab1} immediately above the bump luminosity. 
Both quantities remain zero until we reach 
the vicinity of the HBS, where the $\mu$-profile has the depression caused by the reaction $^{3}$He($^{3}$He,2p)$^{4}$He.
There, the $\mu$ ratio drops vertically because there are no changes of the O abundance yet. When we move further to the right into
the HBS, the H burning in the CNO cycle increases both $\Delta\mu$ and $\Delta X(\mbox{O})$. Fig.~\ref{fig:dodmu} shows
that the $\Delta\mu/\mu_\mathrm{BCE}$ ratio increases faster in the extreme population RGB model, which means that its chemical
structure does not facilitate the penetration of extra mixing deeper into the HBS and dredge up more material with a deficit in O, 
as compared to the normal population RGB model. 
Moreover, in order to attain the same level of the surface O depletion, if it is required by observations,
the extreme population RGB star must have more powerful extra mixing, e.g. if the RGB extra mixing
is driven by rotation then the extreme population stars in GCs must rotate faster than their normal population counterparts
by some reason, which is difficult to understand. 

In the hypothesis that proposes the supermassive MS stars with $M > 10^4\,M_\odot$ as the source of the primordial abundance
variations in GCs, it is sufficient to assume that some low-mass stars in GCs were formed out of more than
90\%  of the material lost by these supermassive stars (Fig.~\ref{fig:ona_mg}A). We also note that
the total CNO abundances in the M\,13 RGB stars measured by \cite{cohen:05}, namely
the [(C+N+O)/H] ratios between -1.4 and -1.1 with the average value -1.23, are very close to those in
our Table~\ref{tab:tab1}. 

Like massive MS stars, the supermassive stars destroy Li very quickly. 
That could be a problem for a hypothesis invoking such stars
as the source of the primordial abundance variations in GCs if low-mass stars with low O and high Li abundances were found
in a GC. The most likely pollution source in that case would be massive AGB stars, because
they produce Li via the convective $^7Be$-mechanism \citep{cameron:71}, while destroying O in the HBB
\citep[e.g.,][]{ventura:10}.
However, there are no stars found in GCs yet that are O-poor and Li-rich at the same time. On the contrary, there are spectroscopic
observations of MS turn-off stars in 47\,Tuc \citep{dobrovolskas:14} and NGC\,6752 \citep{shen:10}, the latter having
[Fe/H] close to that of M\,13, that reveal O-Li correlations, as if material with depleted abundances of both O and Li 
had been mixed with the pristine gas in these GCs. Although \cite{shen:10} claim that the characteristic of their obtained 
O-Li correlation for NGC\,6752 MS turn-off stars requires that the polluting gas had been enriched in Li, 
we notice that a large number of their measured Li abundances have values that are nearly $0.2$\,--\,$0.3\,\mathrm{dex}$ higher 
than those of both the Spite plateau Li abundance \citep{spite:82} and Li abundances in MS turn-off stars from 
the same GC reported 
by \cite{gruyters:14}.
After their reduction by $0.2$\,--\,$0.3\,\mathrm{dex}$, the (logarithmic) Li abundances from \cite{shen:10} will 
probably show a one-to-one correlation with [O/Fe], as predicted by mixing of pristine gas with material from massive 
and supermassive MS stars.

\begin{figure}
\includegraphics[width=8cm,bb = 76 182 493 495]{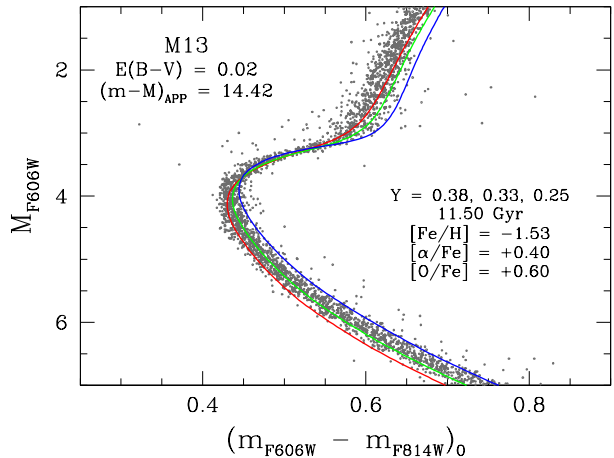}
\caption{\small The three 11.5 Gyr isochrones generated with the Victoria stellar evolution code for the indicated combinations of
         $Y$, [Fe/H] and [$\alpha$/Fe] that are close to those assumed in our normal ($Y=0.25$, blue), 
         intermediate ($Y=0.33$, green) and extreme ($Y=0.38$, red) compositions of 
         low-mass stars in M\,13 are compared with the M\,13 HST ACS color-magnitude diagram.
         }
\label{fig:m13isohbacs}
\end{figure}

\begin{figure}
\includegraphics[width=8cm,bb = 105 244 482 531]{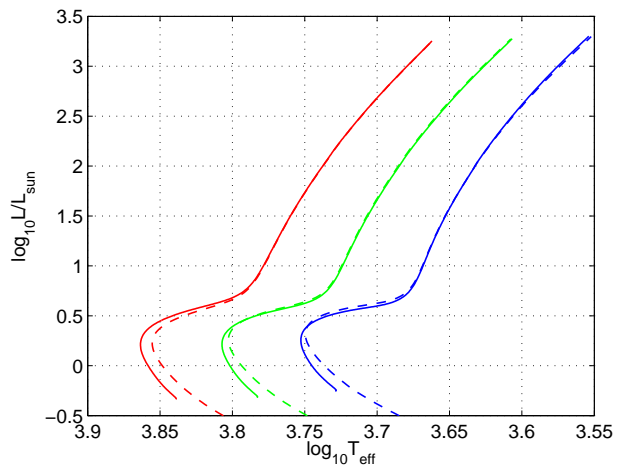}
\caption{\small The isochrones (solid curves) for the normal (blue), intermediate (green) and extreme (red) populations of
         low-mass stars in M\,13 and the stellar evolutionary tracks (dashed curves) of stars with the masses 
         $0.8\,M_\odot$ (blue), $0.7\,M_\odot$ (green) and $0.65\,M_\odot$ (red) that have the corresponding initial compositions 
         from Table~\ref{tab:tab1} and matching red gian branches. The blue and red curves have been shifted by
         $\Delta\log_{10}T_\mathrm{eff}=\pm 0.05$ relative to the green curves.
         }
\label{fig:don_trk_iso}
\end{figure}

\section{The evolutionary abundance variations in M13 RGB stars}

We have used the Victoria stellar evolution code to generate three 11.5 Gyr isochrones for the combinations of the initial He mass 
fraction, metallicity and $\alpha$-element enhancement that are close to those assumed for the extreme, intermediate and 
normal populations of low-mass stars in M13 (Table~\ref{tab:tab1}). These isochrones are transformed to the HST ACS 
photometric system (using the colour--$T_{\rm eff}$ relations given by \citealt{casagrande:14})
and compared with the HST CMD data for M13 in Fig.~\ref{fig:m13isohbacs}, for which we have selected 
the 100 stars in each 0.10 mag bin that have the smallest error bars on the observed magnitudes and colors.
The RGB segments of the isochrones appear to be redder than their corresponding observed CMD. A number of possible
sources of this discrepancy are discussed in Section 6.1.2 of the paper of \cite{vandenberg:13}. It is also possible that
a majority of M13 stars have initial chemical composition that is actually closer to the intermediate one, with $Y\approx 0.33$, 
as suggested by the O-Na anticorrelation in Fig.~\ref{fig:ona_mg}A. In this case, the fiducial theoretical isochrone for M13 
should be the green one.

The individual evolutionary tracks of stars with the masses $0.65\,M_\odot$, $0.7\,M_\odot$ and $0.8\,M_\odot$ calculated for
the first, second and third initial compositions, respectively, are found to have RGBs coinciding with the RGBs of their 
corresponding isochrones (Fig.~\ref{fig:don_trk_iso}). Therefore, we have chosen these masses as the initial ones for our study of
the evolutionary abundance variations, caused by the RGB extra mixing, in the M13 low-mass stars belonging to the extreme,
intermediate and normal populations, using for them the initial abundances from Table~\ref{tab:tab1}. In this study, we employ
the revision 3251 of MESA instead of the Victoria code because the latter cannot model extra mixing on the upper RGB, 
although the Victoria code produces better isochrones than MESA because it accounts for the atomic diffusion and its counteracting 
turbulent mixing on the MS, which has not yet been implemented in the MESA code.

\begin{figure*}
\includegraphics[width=15cm,bb = 103 257 544 603]{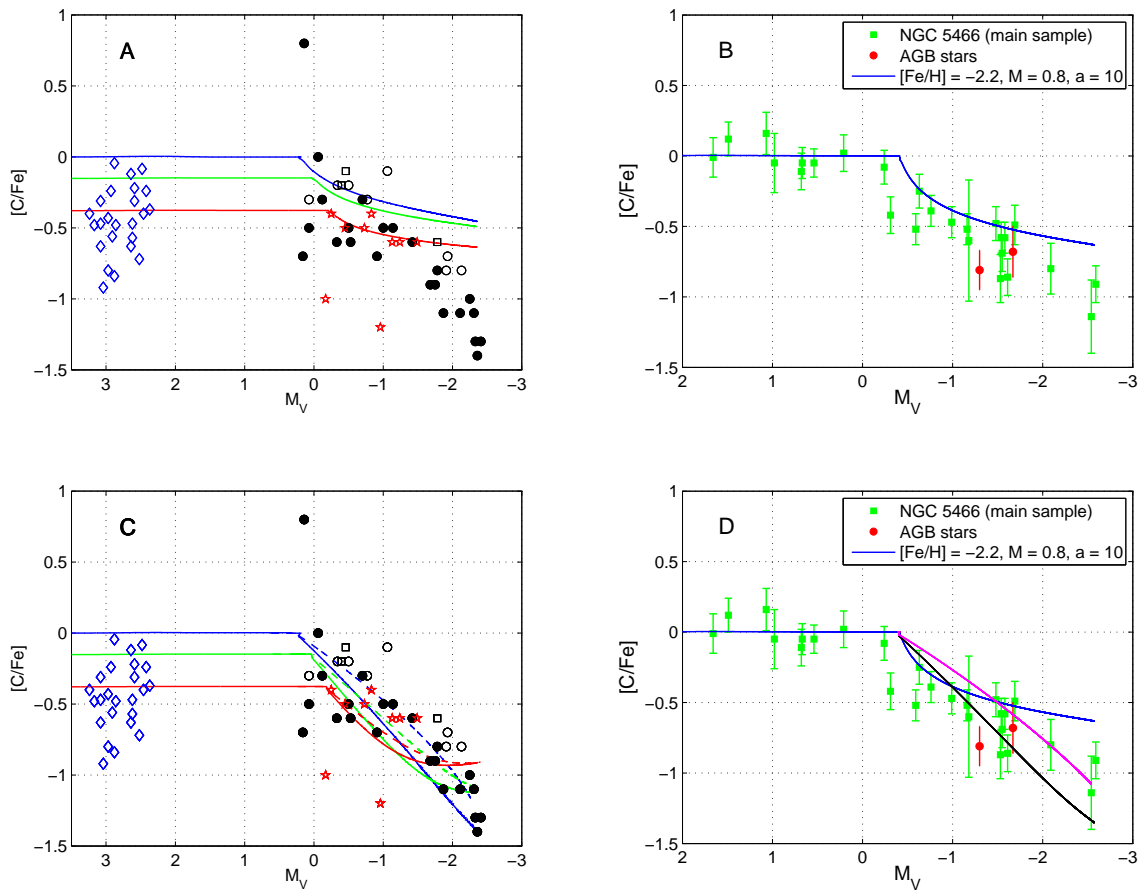}
\caption{\small Panel A: the evolution of the surface C abundance in the stars with the masses $0.8\,M_\odot$ (blue),
         $0.7\,M_\odot$ (green) and $0.65\,M_\odot$ (red) and initial chemical compositions from the second, third and fourth 
         columns of Table~\protect\ref{tab:tab1}, respectively, calculated from the MS to the RGB tip. The RGB extra mixing is
         modeled using the thermohaline diffusion coefficient (equation 25) from \protect\cite{denissenkov:10} 
         with the salt-finger aspect 
         ratio $a=7$ that produces the maximum possible decrease of [C/Fe]. The results of these calculations are compared with
         the observational data for M\,13 MS turn-off and subgiant stars (the open blue diamonds) and RGB stars (the other
         symbols, except the red star symbols which represent AGB stars) from \protect\cite{briley:02}. 
         Panel B: the same as in panel A, but for the stars of the single stellar population GC
         NGC5466 observed by \protect\cite{shetrone:10} and $a=10$. Panel D: The same as in panel B, but
         here we have also used the parametric prescription for the RGB extra mixing with the mixing depth 
         $\log_{10}(r_\mathrm{mix}/R_\odot) = -1.35$ (same as for the thermohaline mixing) and diffusion coefficient 
         $D_\mathrm{mix} = \alpha K$, where $K$ is the radiative diffusivity and $\alpha = 0.02$ (magenta line) and 0.03 
         (black line).  Panel C: the same as in panel D, but for the M\,13 stars with the dashed and solid curves representing 
         the cases of $\alpha = 0.02$ and 0.03.
         }
\label{fig:m13_NGC5466_abu}
\end{figure*}

The red, green and blue curves in Fig.~\ref{fig:m13_NGC5466_abu}A show the evolution of the surface C abundance 
in the models representing the three populations of low-mass stars in M13,
in which the RGB extra mixing has been modeled using the thermohaline diffusion coefficient (equation 25) 
from \cite{denissenkov:10}.
We have used the salt-finger aspect ratio $a=7$ that gives the maximum possible depletion of [C/Fe] in these models.
The open blue diamonds in Fig.~\ref{fig:m13_NGC5466_abu}A are M13 MS turn-off and subgiant stars for which the [C/Fe] values
were determined by \cite{briley:02}. They demonstrate a pattern characteristic of the equilibrium CNO cycle --- the floor
at [C/Fe]\,$\approx -0.8$ \citep{denissenkov:98}, which supports the idea that the primordial abundance variations in GC stars
were produced in H burning at a sufficiently high temperature for the CNO cycle to reach equilibrium. The rest of the symbols
in Fig.~\ref{fig:m13_NGC5466_abu}A, except the red star symbols that represent low-mass AGB stars, present the [C/Fe] data 
for RGB stars with $M_V < +0.8$ compiled by \cite{smith:06} from the literature. We have used $(m-M)_V = 14.42$ as the distance 
modulus for M13 and applied the correction $\Delta$[C/Fe]\,$=+0.4$ to all of the RGB [C/Fe] values, 
as recommended in the cited paper.
We see that the thermohaline convection driven by the $^{3}$He burning produces shallow evolutionary declines of
[C/Fe] incompatible with the observational data for the M13 RGB stars. 

From Fig.~\ref{fig:m13_NGC5466_abu}B, the same conclusion can be made for the NGC\,5466 stars studied by \cite{shetrone:10}, 
for which the value of $a=10$ gives
a maximum effect for the evolutionary depletion of [C/Fe] on the upper RGB (our defined maximal-mixing salt-finger
aspect ratio slightly depends on the metallicity). This GC is unique because it does not appear
to have any primordial abundance variations \citep{shetrone:10}. This conclusion has recently been supported by the new HST
photometric data extended to include UV passbands, according to which NGC\,5466 does not appear to have multiple CMDs 
\citep{piotto:14}.
Therefore, it can be used to calibrate the depth and rate of the RGB extra 
mixing. To eliminate the mixing depth as a free parameter, we assume that it is equal to the almost universal depth that we usually
find for the $^3$He-driven thermohaline convection in upper RGB models of different metallicities, 
i.e. $\log_{10}(r_\mathrm{mix}/R_\odot) = -1.35$. This depth guarantees that only 
the products of H burning in the CN branch of the CNO cycle are dredged up from the HBS to BCE, as indicated by observations. 
In the absence of a good candidate for the mechanism of the RGB extra mixing, we assume that it can be modeled as a diffusion 
process with a diffusion coefficient $D_\mathrm{mix}$ proportional to the radiative diffusivity
\bea
K = \frac{4acT^3}{3\kappa C_P\rho^2},
\eea
where $a$ is the radiation constant, $c$ the speed of light, $\kappa$ is the Rosseland mean opacity, $C_P$ is the specific
heat at constant pressure and $\rho$ is the density. This assumption makes sense as long as the RGB extra mixing operates on
a thermal timescale, when the radiative heat diffusion facilitates it by reducing temperature contrasts between
rising and sinking fluid parcels. In Fig.~\ref{fig:m13_NGC5466_abu}D, the magenta and black curves are
obtained with $D_\mathrm{mix} = \alpha K$ for $\alpha = 0.02$ and 0.03, respectively. When we employ the same parameters
of the RGB extra mixing in models of the M13 low-mass stars, we get very good agreement with the observational data 
(Fig.~\ref{fig:m13_NGC5466_abu}C), in spite of the fact that the two GCs have different metallicities, [Fe/H]\,$=-1.53$ for M13 and
[Fe/H]\,$=-2.2$ for NGC\,5466. 

\section{CMDs for the three populations of low-mass stars in M13}
\label{sec:CMDs}

The color difference between our theoretical isochrones for the normal and extreme populations of low-mass stars in M13
in Fig.~\ref{fig:m13isohbacs} is approximately as large as the width of its CMD observed with the HST ACS. 
Therefore, our assumption that M13 has the populations
of low-mass stars with $Y$ varying between 0.25 and 0.38 cannot be rejected on the basis of 
these photometric data\footnote{The implications of high helium abundances and/or a wide range in $Y$ for the horizontal
branch of M13 will be considered in a separate paper (P.~Denissenkov et al., in preparation).  
Some studies \protect\citep[e.g.][]{catelan:09} have argued in support of a normal helium content for this cluster, while
others have estimated $Y_\mathrm{max}\approx 0.31$ for M13 HB stars \protect\citep[e.g.][]{dalessandro:13}.}.
This conclusion appears even more true when we take into consideration that both the extreme and normal populations are 
likely to be poorly presented in M13, as
compared to its intermediate population. This possibility is supported by the fact that
the O-Na anti-correlation for M13 in Fig.~\ref{fig:ona_mg}A  includes 63\% of the intermediate-population RGB stars, 
while the normal and extreme populations contribute only 15\% and 22\% \citep{johnson:12}. Also, the M13 RGB stars with 
the most extreme abundance anomalies are predominantly
located near the RGB tip, where the three isochrones almost converge. 

We remind the reader that our subdivision of the M13 low-mass stars into the three distinct populations is 
an approximation that has been made
using the rather arbitrary selection criteria. In fact, we assume that variations of abundances of the p-capture 
elements in the material out of which low-mass stars formed in GCs were smooth, reflecting dilution of the GC pristine gas
by gas partially processed in supermassive MS stars, with a random but continuously varying fraction of the latter. 
This assumption seems to contradict to multiple CMDs found
in a constantly increasing number of GCs \citep[e.g.,][]{monelli:13,piotto:14}. Except a small number of GCs with intrinsic
and almost discrete variations of [Fe/H] (e.g., $\omega$\,Cen and M22) or [(C+N+O)/Fe] (NGC2808), which can only be explained by
several star formation episodes, multiple CMDs in most other GCs with nearly constant [Fe/H] and [(C+N+O)/Fe]
values in each of them can probably be explained by the fact that their narrow-band photometric observations filter out
some special spectral features characteristic of the p-capture element abundance anomalies. 
At least, this seems to be true when one
uses the new photometric index $c_{U,B,I} = (U-B) - (B-I)$ that has recently been proposed by \cite{monelli:13} to help reveal
multiple stellar populations in GCs more easily. Indeed, as \cite{sbordone:11} have shown, the enhanced abundance of N 
in the population of stars polluted by the
products of H burning (the intermediate and primordial populations in our case) leads to a significant increase of the $U$
magnitude because of the higher concentrations of the CN and NH molecules in these stars that absorb more light in the $U$ band.
From our point of view, this is like looking at a rainbow, that has a smooth transition of color from red to violet, through an
anaglyph glasses and, as a result, seeing discrete red and cyan stripes. Our hypothesis is indirectly supported by 
\cite{bastian:13b} who have not found evidence for ongoing star formation within any of 130 Galactic and extragalactic young 
massive clusters surveyed by them.

\begin{figure}
\includegraphics[width=8cm,bb = 114 248 471 531]{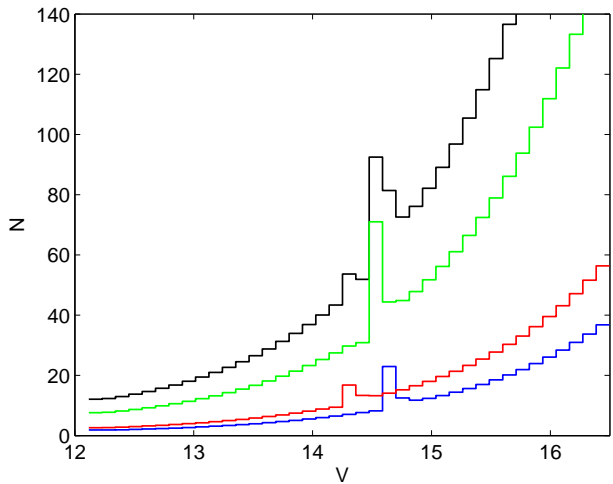}
\caption{\small The theoretical differential luminosity functions for the evolutionary tracks 
        from Fig.~\protect\ref{fig:m13_NGC5466_abu}A
        (the blue, green and red histograms) constructed assuming that the three populations contribute 15\%, 63\% and 22\% 
        to the total population of stars in M\,13, like in its O-Na anti-correlation \protect\citep{johnson:12}, as well as their
        composite luminosity function (the black histogram).
         }
\label{fig:m13_lf}
\end{figure}

From Fig.~\ref{fig:m13_NGC5466_abu}C, it is seen that the evolutionary tracks of the low-mass stars belonging to the different
populations in M13 have slightly different bump luminosities, at which the RGB extra mixing starts to operate, 
increasing with $Y$. A differential luminosity function constructed
for the normal population (the blue curve in Fig.~\ref{fig:m13_lf}) has a bump at $V\approx 14.7$ which is very close to
its observed location in M13, $V\approx 14.75$, as reported by \cite{sandquist:10}, or $V\approx 14.77$, according to the most
recent HST GC data analysis by \cite{nataf:13}. However, if the M13 low-mass stars represent 
a mixture of the three populations, then location and width of the bump depend on its relative strength and a number of stars
in the different populations. It turns out that the former decreases when $Y$ increases.
The red, green and blue curves in Fig.~\ref{fig:m13_lf} show the differential luminosity functions for the M13 three 
populations constructed assuming that they include 22\%, 63\% and 15\% of all the M13 stars, like in its O-Na anti-correlation, 
while the black curve is their superposition. Given that the intermediate population is dominating in this mixture, the composite 
bump luminosity has shifted to $V\approx 14.6$. This analysis demonstrates that the observed location of the bump luminosity in a GC
with multiple stellar populations can in principle be used to estimate their relative weights.

\section{Conclusion}
\label{sec:concl}

In this paper, we have elaborated on the hypothesis, recently proposed by \cite{denissenkov:14}, that the primordial abundance
anomalies of the p-capture elements and their isotopes in GC stars had been produced during a short time
($\sim 10^5$ years) of H burning in the fully convective supermassive MS stars with $M>10^4\,M_\odot$. Because such stars
are supported against the force of gravity almost entirely by the radiation pressure, they are subject to the diffusive mode of
the Jeans instability, which develops on all length scales and on a timescale comparable to the lifetime of the supermassive stars
\citep{thompson:08,denissenkov:14}. Therefore, those stars might have fallen apart (fragmented) soon after they had formed,
by a moment when only a small fraction of H was transformed into He, as we have assumed. 
It is out of the scope of the present paper 
to discuss how the supermassive stars had formed in the young GCs 
\citep[the two possible formation scenarios are briefly reviewed by][]{denissenkov:14}, 
or how they had lost their mass and how that mass, polluted with the products of H burning,
had been mixed with the GC pristine gas before the low-mass stars formed out of that mixture. Here, we have focused 
on the nucleosynthesis part of the solution.

We have shown that among the massive star candidates for the origin of the primordial abundance anomalies in GCs
proposed so far, such as the massive MS stars (rapidly rotating, members of binary systems, or as massive as $\sim 10^3\,M_\odot$),
massive AGB stars and supermassive MS stars with $M>10^4\,M_\odot$, only the latter have the right temperatures of H burning 
in the range $74\times 10^6\ \mbox{K}\la T_\mathrm{c}\la 78\times 10^6$ K for the successful reproduction of 
the (anti-)correlations between the Al and Mg isotopic abundances, that have now been found to be common for 5 GCs.

The agreement between the primordial abundance anomalies of the p-capture elements in GCs and their corresponding
abundance variations in the mixtures composed of the pristine GC material and H-burning yields
from the supermassive MS stars with $M>10^4\,M_\odot$ is so good that it is worth trying to (1) use such mixtures, e.g. those from
Table~\ref{tab:tab1}, as the initial compositions for low-mass star models that are supposed to belong to different populations of
stars in a GC with multiple stellar population, (2) allow these models to evolve and (3) see how various physical assumptions, 
e.g. the parameters of the RGB
extra mixing and mass loss, will affect their surface chemical composition and evolution on the CMD in comparison with
observational data. We have tried this for the GC M\,13 (NGC6205) and found that the evolutionary decline of
the C abundance in its upper RGB stars cannot be explained by the $^3$He-driven thermohaline convection. We have estimated
the depth and rate of the RGB extra mixing that allow to reproduce the observational decrease of [C/Fe] with $M_V$. 
They turn out to have
the same values for both M\,13 that has the most extreme primordial abundance anomalies and NGC5466 that does not have such 
anomalies. The fact that the two GCs also have different metallicities supports the old idea about the universality of the RGB
extra mixing \citep[e.g.][]{denvdb:03}. However, we are still missing the understanding of its physical mechanism.
Given that the supermassive stars considered by \cite{denissenkov:14} are hypothetical objects and that neither the massive MS
stars nor the massive AGB stars can reproduce the (anti-)correlations between the Al and Mg isotopic abundances in GCs,
the source of the primordial abundance variations in GCs also remains uncertain. This leaves us with a problem
with two unknowns.

\section*{Acknowledgments} 

This research has been supported by the National
Science Foundation under grants PHY 11-25915 and AST 11-09174. This
project was also supported by JINA (NSF grant PHY 08-22648). 
Falk Herwig and Don VandenBerg acknowledge funding from Natural Sciences and Engineering
Research Council of Canada.
Pavel Denissenkov is grateful to the staff of the Max-Planck-Institut f\"{u}r Astrophysik,
where this project was completed, for their warm hospitality.
We thank M. Salaris who has generously provided us with the Sbordone et al. bolometric corrections.

\bibliography{paper}

\label{lastpage}

\end{document}